\documentclass[12pt]{article}
\usepackage{epsf}
\usepackage{axodraw}
\usepackage{cite}
\newcommand{\real}{{\rm Re}\,} 
\newcommand{\beq}{\begin{equation}}
\newcommand{\eeq}{\end{equation}}
\newcommand{\beqn}{\begin{eqnarray}}
\newcommand{\eeqn}{\end{eqnarray}}
\newcommand{\bea}{\begin{eqnarray}}
\newcommand{\eea}{\end{eqnarray}} 
\newcommand{\be}{\begin{equation}}
\newcommand{\ee}{\end{equation}} 

\begin{document}
\begin{titlepage}
\begin{flushright}MADPH-98-1073
\end{flushright}
\begin{flushright}FSU-HEP-980812
\end{flushright}
\begin{flushright}BNL-HET-98/27 
\end{flushright} 
\begin{flushright}August, 1998 
\end{flushright}
\vspace{2truecm}
\begin{center}
{\large\bf
QCD Corrections to  Associated Higgs Boson-Heavy Quark 
Production}
\\
\vspace{1in}
{\bf S.~Dawson}\\
{\it Physics Department, Brookhaven National 
Laboratory,\\
Upton, NY 11973, USA}
\\ 
\vspace{.25in}
{\bf L.~Reina}\\
{\it  
Physics Department, 
University of Wisconsin, \\
Madison, WI 53706, USA\\
and\\ 
Physics Department, Florida State University, \\
Tallahassee, FL 32306, USA}
\vspace{1in}  
\end{center}
\begin{abstract} 
We compute the ${\cal O}(\alpha_s)$ QCD corrections to the inclusive
process $e^+e^-\rightarrow t \bar t h$.  Although the total rate is
small, it has a distinctive experimental signature and can potentially
be used to measure the top quark-Higgs boson Yukawa coupling.  The QCD
corrections increase the rate by a factor of roughly $1.5$ for
$e^+e^-\rightarrow t \bar t h$ at $\sqrt{s}\!=\!500~\mbox{GeV}$ and
$M_h=100~\mbox{GeV}$. At $\sqrt{s}=1~\mbox{TeV}$, the corrections are
small.  
\end{abstract}
\end{titlepage}
\clearpage 

\section{Introduction}
\label{sec:intro}

The search for the Higgs boson is one of the most important objectives
of present and future colliders.  A Higgs boson or some object like it
is needed in order to give the $W^\pm$ and $Z$ gauge bosons their
observed masses and to cancel the divergences which arise when
radiative corrections to electroweak observables are computed.  We
have few clues as to the expected mass of the Higgs boson, which ${\it
a~priori}$ is a free parameter of the theory. Direct experimental
searches for the Standard Model Higgs boson at LEP and LEP2 yield the
limit,~\cite{leplim}

\beq  
M_h>89.8\,\,\mbox{GeV}\,\,\,\mbox{at 95\% c.l.}
\eeq

\noindent LEP2 will eventually extend this limit to 
around $M_h>109\,\,\mbox{GeV}\,\,\mbox{at 95\%}$ c.l.  Also, from the
analysis of the electroweak precision measurements an indirect upper
bound can be found,~\cite{leplim}

\beq
M_h<280\,\,\mbox{GeV}\,\,\,
\mbox{at 95\% c.l.}
\,\,\,\,
\eeq 

Above the LEP2 direct search limit and below the $2Z$ boson pair
production threshold is termed the \emph{intermediate mass region} and
is the most difficult Higgs mass region to probe experimentally.  In
this mass range the associated production of a Higgs boson with either
a gauge boson or a fermion - antifermion pair can be an important
discovery channel.  In this paper, we focus on the associated
production of a Higgs boson with a heavy quark pair.

The associated production of a Higgs boson with a top quark pair in
$e^+e^-$ collisions has a small rate, around $1~fb$ for
$\sqrt{s}=500~\mbox{GeV}$ and $M_h\sim 100~\mbox{GeV}$.  However, the
signature, $e^+e^-\rightarrow t {\overline t}h
\rightarrow W^+W^- b {\overline b} b {\overline b}$, is 
distinctive.~\cite{desystud} The experimental viability of this
signature has not yet been carefully evaluated.

Once the Higgs boson has been discovered, it will be important to
measure its couplings to fermions and gauge bosons.  These couplings
are completely determined in the Standard Model with no adjustable
parameters.  The couplings to the gauge bosons can be measured through
the associated production processes, $e^+e^-\rightarrow Zh$, $ q
{\overline q}^\prime \rightarrow W^+h$, and $ q {\overline
q}\rightarrow Z h$, and through vector boson fusion,
$W^+W^-\rightarrow h$ and $ZZ\rightarrow h$.  The couplings of the
Higgs boson to fermions are more difficult to measure,
however.~\cite{snow}

The process $e^+e^-\rightarrow t {\overline t} h$ provides a direct
mechanism for measuring the $t {\overline t} h$ Yukawa coupling.
Since this coupling can be significantly different in a supersymmetric
model from that in the Standard Model, the measurement would provide a
mean of discriminating between models.  The $ t {\overline t}h$ Yukawa
coupling also enters into the rates for $gg\rightarrow h$ and
$h\rightarrow \gamma\gamma$, as these processes have large
contributions from top quark loops.  However, in these cases it is
possible that there is unknown new physics which also enters into the
rate and dilutes the interpretation of the signal as the measurement
of the $t {\overline t } h$ coupling.  Ref.~\cite{snow} estimates that
for a Higgs boson with mass less than $100~\mbox{GeV}$, both the $t
{\overline t}h$ and $b {\overline b}h$ couplings can be measured to an
accuracy of roughly $\pm 30~\%$ at the LHC with $600~fb^{-1}$ of data.
It is possible that a high energy lepton collider could obtain a
higher precision on the measurement of the couplings.~\cite{snow}

The QCD corrections to the associated production of a Higgs boson with
a heavy quark pair are the subject of this paper.  They have been
independently computed by Dittmaier ${\it et. al.}~$~\cite{dittqcd} We
compute the QCD corrections to the process $e^+e^-\rightarrow t\bar t
h$, including only the photon exchange diagrams, which constitute the
dominant contribution to the cross section.  Our corrections are valid
for all values of the Higgs boson mass and for any energy.
Section~\ref{sec:lowestorder} contains a review of the lowest order
rate, including a discussion of the relative importance of the $Z$
boson contribution (which we neglect) and the Higgs bremsstrahlung
from the $Z$ boson.  The Higgs bremsstrahlung from the $Z$ is always
less than a few percent effect and so the $e^+e^-\rightarrow t\bar t
h$ process remains a candidate for measuring the coupling of the Higgs
to the top quark.
  
The ${\cal O}(\alpha_s)$ contributions are discussed in
Sections~\ref{sec:real} and \ref{sec:virtual}.  The contributions from
the gluon bremsstrahlung diagrams and the separation of the
calculation of the gluon emission into hard and soft pieces are
presented in Section~\ref{sec:real}.  The one-loop virtual
contributions are discussed in Section ~\ref{sec:virtual}. Our
numerical results for $\sqrt{s}=500~\mbox{GeV}$ and $1~\mbox{TeV}$ are
presented in Section~\ref{sec:results}.

The corrections to $e^+e^-\rightarrow t {\overline t} h$ have
previously been estimated in an approximation valid at high energy and
for $M_h << M_t$.~\cite{dr}.  In the region where the approximation is
valid, $\sqrt{s} > 1~\mbox{TeV}$ and $M_h<100~\mbox{GeV}$, the QCD
corrections were estimated to be small.  We end with a comparison of
our results with the approximate calculation.

\section{$e^+e^-\rightarrow t {\overline t} h$: Lowest 
Order }
\label{sec:lowestorder}
  
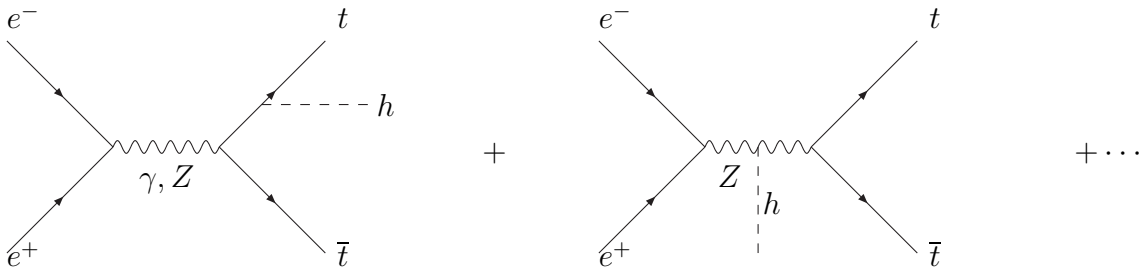
\begin{figure}[hbtp]
\begin{picture}(100,100)(-20,-5)
\SetScale{0.8}
\ArrowLine(0,100)(50,50)
\ArrowLine(0,0)(50,50)
\Photon(50,50)(100,50){3}{6}
\ArrowLine(100,50)(150,100) 
\ArrowLine(100,50)(150,0)
\DashLine(120,70)(170,70){5}
\put(180,35){$+$} 
\put(50,25){$\gamma, Z$} 
\put(0,85){$e^-$}
\put(0,-5){$e^+$}
\put(140,52){$h$}
\put(125,85){$t$}
\put(125,-5){${\overline t}$}
\SetScale{1}
\end{picture}
\begin{picture}(100,100)(-140,-5)
\SetScale{0.8}
\ArrowLine(0,100)(50,50)
\ArrowLine(0,0)(50,50)
\Photon(50,50)(100,50){3}{6}
\DashLine(75,50)(75,0){5}
\ArrowLine(100,50)(150,100) 
\ArrowLine(100,50)(150,0)
\put(180,35){$+ \cdots$} 
\put(45,25){$Z$} 
\put(0,85){$e^-$}
\put(0,-5){$e^+$}
\put(62,15){$h$}
\put(125,85){$t$}
\put(125,-5){${\overline t}$}
\SetScale{1}
\end{picture} 
\caption[]{Feynman diagrams contributing to the lowest
order process, $e^+e^-\rightarrow t {\overline t} h$.}
\label{lofeyndiag}
\end{figure}

The cross section for $e^+e^-\rightarrow t\bar t h$ occurs through the
Feynman diagrams of Fig. \ref{lofeyndiag} and was first calculated in
\cite{gounaris} (photon-exchange contribution only) and 
then completed in \cite{djouadi} (both photon and $Z$-exchange
contributions).  We write the cross section in the form

\begin{figure}[bt]
\centering
\epsfxsize=4.5in
\leavevmode\epsffile{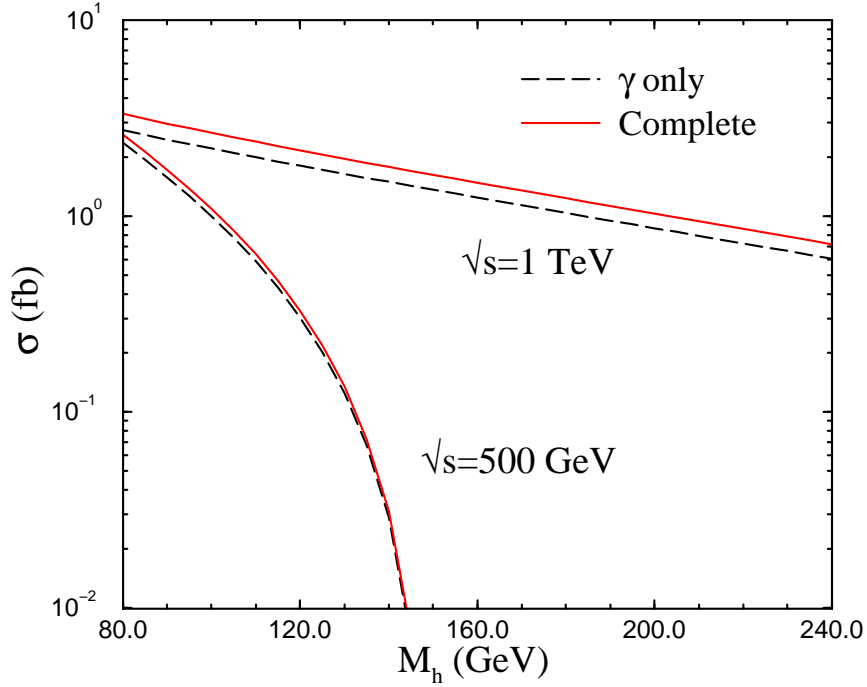}
\caption[]{Lowest order cross section for 
$e^+e^-\rightarrow t {\overline t}h$ at $\sqrt{s}=500$~GeV and
$\sqrt{s}=1~\mbox{TeV}$.  The curve labelled {\it complete} includes
both $\gamma$ and $Z$ exchange, along with bremsstrahlung from the $Z$ 
boson. We take $M_t=175$~GeV.}
\label{tthlofig}
\end{figure}

\bea 
\frac{d\sigma(e^+e^-\rightarrow t\bar t h^0)}{dx_h} &=& 
N_c\frac{\sigma_0}{(4\pi)^2}\left\{
\left[Q_e^2Q_t^2+\frac{2Q_eQ_tV_eV_t}{1-M_Z^2/s}+
\frac{(V_e^2+A_e^2)(V_t^2+A_t^2)}{(1-M_Z^2/s)^2}\right]G_
1\right.
\nonumber\\
&&\left.+ 
\frac{V_e^2+A_e^2}{(1-M_Z^2/s)^2}\left[A_t^2\sum_{i=2}^6G
_i+
V_t^2(G_4+G_6)\right]+\frac{Q_eQ_tV_eV_t}{1-M_Z^2/s}G_6 
\right\}
\,\,\,\,,
\label{dsig0}
\eea

\noindent where $\sigma_0\!=\!4\pi\alpha^2/3s$, $\alpha$ 
is the QED fine structure constant, $N_c=3$ is the number of colors,
$x_h=2 E_h/\sqrt{s}$ with $E_h$ the Higgs boson energy, and $Q_i$,
$V_i$ and $A_i$ ($i\!=\!e$, $t$) denote the electromagnetic and weak
couplings of the electron and of the top quark respectively,

\be
V_i 
=\frac{2I_{3L}^i-4Q_is_W^2}{4s_Wc_W}\,\,\,\,\,\,\,\,,\,\,
\,\,\,\,\
A_i = \frac{2 I_{3L}^i}{4s_Wc_W}\,\,\,\,,
\ee

\noindent with $I_{3L}^i\!=\!\pm 1/2$ being the weak 
isospin of the left-handed fermions and $s_W^2\!=\!1-c_W^2=0.23$.
(The contribution from the photon alone can be trivially found by
setting $V_i=A_i=0$).

The coefficients $G_1$ and $G_2$ describe the radiation of the Higgs
boson off the top quark (both photon and $Z$ boson exchange) and are
given by,

\bea
\label{th_coeff}
G_1&=&\frac{2\,g_t^2}{s^2\left(\hat\beta^2 - x_h^2 
\right)x_h}
     \left(\phantom{\frac{1}{2}}\!\!\!\!
        -4\hat\beta\,\left( 4 M_t^2 - M_h^2 \right) \,
        \left( 2 M_t^2 + s \right) x_h + \right.\\
&& \!\!\!\!\!\!\!\!\!\!\!\!\!\!\!\!\!
        \left.\left(\hat\beta^2 - x_h^2 \right) 
        \left( 16M_t^4 + 2M_h^4 - 
          2M_h^2 s x_h + s^2 x_h^2 - 4 M_t^2  
        \left( 3 M_h^2 - 2 s - 2 s x_h \right)  \right) 
\,
        \log 
\left({\frac{x_h+\hat\beta}{x_h-\hat\beta}}\right)
       \right)\,\,\,, \nonumber\\
G_2&=& \frac{-2\,g_t^2}{s^2\,\left(\hat\beta^2 - x_h^2 
\right) x_h}
     \left(\phantom{\frac{1}{2}}\!\!\!\! \hat\beta\, x_h
        \left( -96 M_t^4 + 24 M_t^2 M_h^2 - 
          \left( -M_h^2 + s + s x_h \right) 
           \left( -\hat\beta^2 + x_h^2 \right)  
\right)\right.  +
     \nonumber\\
&& \!\!\!\!\!\!\!\!\!\!\!\!\!\!\!\!\!
        \left.   2\left( \hat\beta^2 - x_h^2 \right) 
        \left( 24 M_t^4 + 2\left( M_h^4 - M_h^2 s x_h 
\right)  + 
          M_t^2\left( -14 M_h^2 + 12 s x_h + s x_h^2 
\right)  \right) 
\log\left(\frac{x_h+\hat\beta}{x_h-\hat\beta}\right)
       \right) \,\,\,,\nonumber
\eea

\noindent where $g_t$~is the top quark Yukawa coupling,

\be
g_t\equiv {M_t\over v}\,\,\,,
\ee

\noindent $v=(\sqrt{2}G_F)^{-1/2}\simeq 246~\mbox{GeV}$, 
and

\be
{\hat 
\beta}=\biggl\{{[x_h^2-(x_h^{min})^2][x_h^{max}-x_h]
\over x_h^{max}-x_h+{4 M_t^2\over 
s}}\biggr\}^{1/2}\,\,\,\,,
\ee 

\noindent with $x_h^{min}=2 M_h/\sqrt{s}$ and 
$x_h^{max}=1-4M_t^2/s
+M_h^2/s$.

\noindent 
The other four coefficients, $G_3,\ldots,G_6$ describe the emission of
a Higgs boson from the $Z$-boson and can be written in the following
form,

\bea
\label{Zh_coeff}
G_3&=& \frac{-2\,\hat\beta g_Z^2 M_t^2}
{M_Z^2 \left( M_h^2 - M_Z^2 + s - s x_h \right)^2}
     \left( 4 M_h^4 + 12 M_Z^4 + 2 M_Z^2 s x_h^2 + 
       s^2\left( -1 + x_h \right) x_h^2 - 
\right.\nonumber\\
   && \left.   M_h^2\left( 8 M_Z^2 + s\left( -4 + 4 x_h + 
x_h^2 \right) 
           \right)  \right)\,\,\,, \nonumber\\
G_4 &=& \frac{\hat\beta g_Z^2 M_Z^2}
 {6\left( M_h^2 - M_Z^2 + s - s x_h \right)^2}
     \left( 48 M_t^2 + 12 M_h^2 - s\left( -24 + 
\hat\beta^2 + 24 x_h - 
          3 x_h^2 \right)  \right)\,\,\,, \\
G_5 &=&
\frac{-4\,g_t\,g_Z\,M_t}
{M_Z\,s\left( -M_h^2 + M_Z^2 + s\left( -1 + x_h 
\right)\right) }
     \left(\phantom{\frac{1}{2}}\!\!\!\! \hat\beta\,s  
\left( 6 M_Z^2 + 
     x_h\left( -M_h^2 - s + s x_h \right)\right) + 
\right.\nonumber\\
  && \left.   2\left( M_h^2\left( M_h^2 - 3M_Z^2 + s - s 
x_h \right)  + 
          M_t^2\left( -4 M_h^2 + 12 M_Z^2 + s x_h^2 
\right)  \right) \,
\log\left({\frac{x_h+\hat\beta}{x_h-\hat\beta}}\right)
       \right)\,\,\,, \nonumber\\
G_6 &=&
\frac{8\,g_t\,g_Z\,M_t\,M_Z}
{s\left( -M_h^2 + M_Z^2 + s\,\left( -1 + x_h 
\right)\right)}
     \left( \hat\beta\,s + \left( 4 M_t^2 - M_h^2 + 2 s - 
s x_h \right) 
\log\left({\frac{x_h+\hat\beta}{x_h-\hat\beta}}\right)
       \right) \nonumber\,\,\,\,,
\eea

\noindent  
where $g_Z$ denotes the coupling of the Higgs boson
to the $Z$ boson,

\be
g_Z\!\equiv\!{M_Z\over v}\,\,\,.
\ee

As already observed in Ref.~\cite{djouadi}, the most relevant
contributions are those in which the Higgs boson is emitted from a top
quark leg, i.e. those proportional to $G_1$ and $G_2$ in
Eq.~(\ref{dsig0}).  The contribution from the Higgs boson coupling to
the $Z$ boson is always less than a few per cent at $\sqrt{s}=500$~GeV
and $1$~\mbox{TeV} and can safely be neglected.  In
Fig.~\ref{tthlofig}, we show the complete cross section for
$e^+e^-\rightarrow t {\overline t}h$ production and also the
contribution from the photon exchange contribution only.  We see that
at both $\sqrt{s}=500$~GeV and $1$~TeV, the cross section is very well
approximated by the photon exchange only.  In the remainder of the
paper, we will consider only the photon exchange contribution,
neglecting the $Z$ boson exchange contribution everywhere.

\section{Real Gluon Emission}
\label{sec:real}

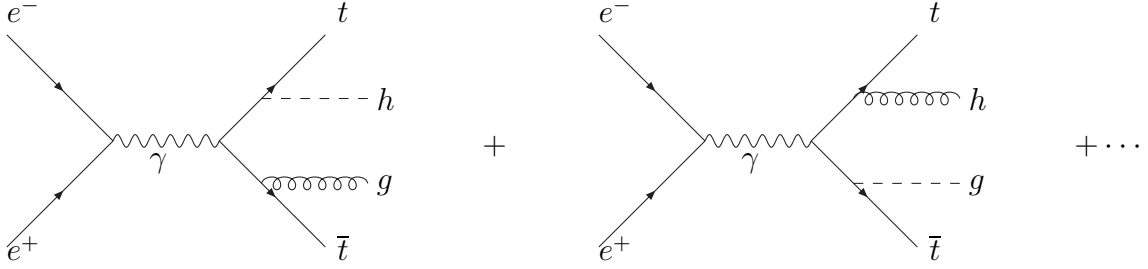
\begin{figure}[btp]
\begin{picture}(100,100)(-20,-5)
\SetScale{0.8}
\ArrowLine(0,100)(50,50)
\ArrowLine(0,0)(50,50)
\Photon(50,50)(100,50){3}{6}
\ArrowLine(100,50)(150,100) 
\ArrowLine(100,50)(150,0)
\DashLine(120,70)(170,70){5}
\Gluon(120,30)(170,30){3}{6} 
\put(180,35){$+$} 
\put(54,30){$\gamma$} 
\put(0,85){$e^-$}
\put(0,-5){$e^+$}
\put(140,52){$h$}
\put(140,22){$g$} 
\put(125,85){$t$}
\put(125,-5){${\overline t } $}
\SetScale{1}
\end{picture}
\begin{picture}(100,100)(-140,-5)
\SetScale{0.8}
\ArrowLine(0,100)(50,50)
\ArrowLine(0,0)(50,50)
\Photon(50,50)(100,50){3}{6}
\ArrowLine(100,50)(150,100) 
\ArrowLine(100,50)(150,0)
\Gluon(120,70)(170,70){3}{6}
\DashLine(120,30)(170,30){5}
\put(180,35){$+ \cdots$} 
\put(54,30){$\gamma$} 
\put(0,85){$e^-$}
\put(0,-5){$e^+$}
\put(140,52){$h$}
\put(140,22){$g$} 
\put(125,85){$t$}
\put(125,-5){${\overline t}$}
\SetScale{1}
\end{picture}
\label{glufig}
\caption[]{Sample Feynman diagrams for the process 
$e^+e^-\rightarrow
t {\overline t}hg$.} 
\end{figure} 

The ${\cal O}(\alpha_s)$ inclusive cross section for
$e^+e^-\rightarrow t {\overline t} h$ receives contributions from real
gluon emission from the final quark legs,

\be
e^+(p)+ e^-(q)\rightarrow t(p_t)+ {\overline 
t}(p_t^\prime)+ 
h(p_h)+g(k)\,\,\,\,,
\ee 

\noindent as shown in Fig.~\ref{glufig}.
The cross section can be separated into hard and soft pieces by
introducing an arbitrary separation $E_{min}$ on the gluon momenta,
such that for $E_g < E_{min}$, we can use the eikonal approximation
and calculate the cross section analytically, while for $E_g>
E_{min}$, we integrate over the phase space numerically.  The final
result is of course independent of the cut-off $E_{min}$ (see
Section~\ref{sec:realhard}). 

\subsection{Soft Gluon Radiation}
\label{sec:realsoft}

For soft gluons, $E_g < E_{min}$, we neglect the momenta of the
radiated gluons everywhere but in the singular propagators.  In the
soft gluon approximation, we find the contribution from the radiated
gluons,

\be
\biggl({d\sigma\over d x_h}\biggr)_{soft}=
-\biggl({d\sigma\over d x_h}\biggr)_0 \biggl({g_s^2 
C_F\over (2 \pi)^3} 
\biggr)
\int_{\mid k\mid < E_{min}}
{d^3 k\over 2 \omega(k)}
\biggl\{{M_t^2\over (p_t\cdot k)^2}
+{M_t^2\over (p_{t^\prime}\cdot k)^2}
-{2 p_t\cdot p_{t^\prime}\over
(p_t\cdot k)(p_{t^\prime}\cdot k)}\biggr\}\,\,\,,
\ee 

\noindent where $k$ is the gluon momentum, $\omega(k)\equiv
\sqrt{\mid{\vec k}\mid^2+m_g^2}$ and we have introduced a small gluon
mass $m_g$ to regulate the infrared divergences occurring in the soft
gluon emission.  The dependence on the gluon mass will be cancelled by
contributions from the virtual graphs which are also evaluated with a
non-zero gluon mass (see Section~\ref{sec:irdiv}).  The lowest order
result can be found from Eq.~(\ref{dsig0}).

\begin{figure}[bt]
\centering
\epsfxsize=4.5in
\leavevmode\epsffile{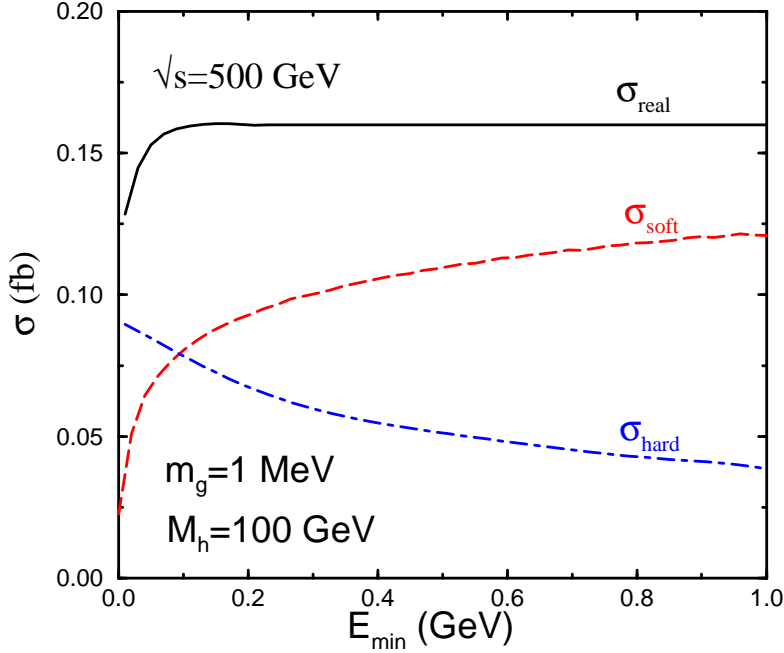}
\caption[]{Contributions  to the
process $e^+e^-\rightarrow t {\overline t} h g$ as a function of the
cut-off on the soft gluon energy, $E_{min}$, for fixed gluon and Higgs
boson masses at $\sqrt{s}=500$~GeV. We take $M_t=175$~GeV and
$\alpha_s(M_t)=.11164$.}
\label{eminfig}
\end{figure}

\begin{figure}[bt]
\centering
\epsfxsize=4.5in
\leavevmode\epsffile{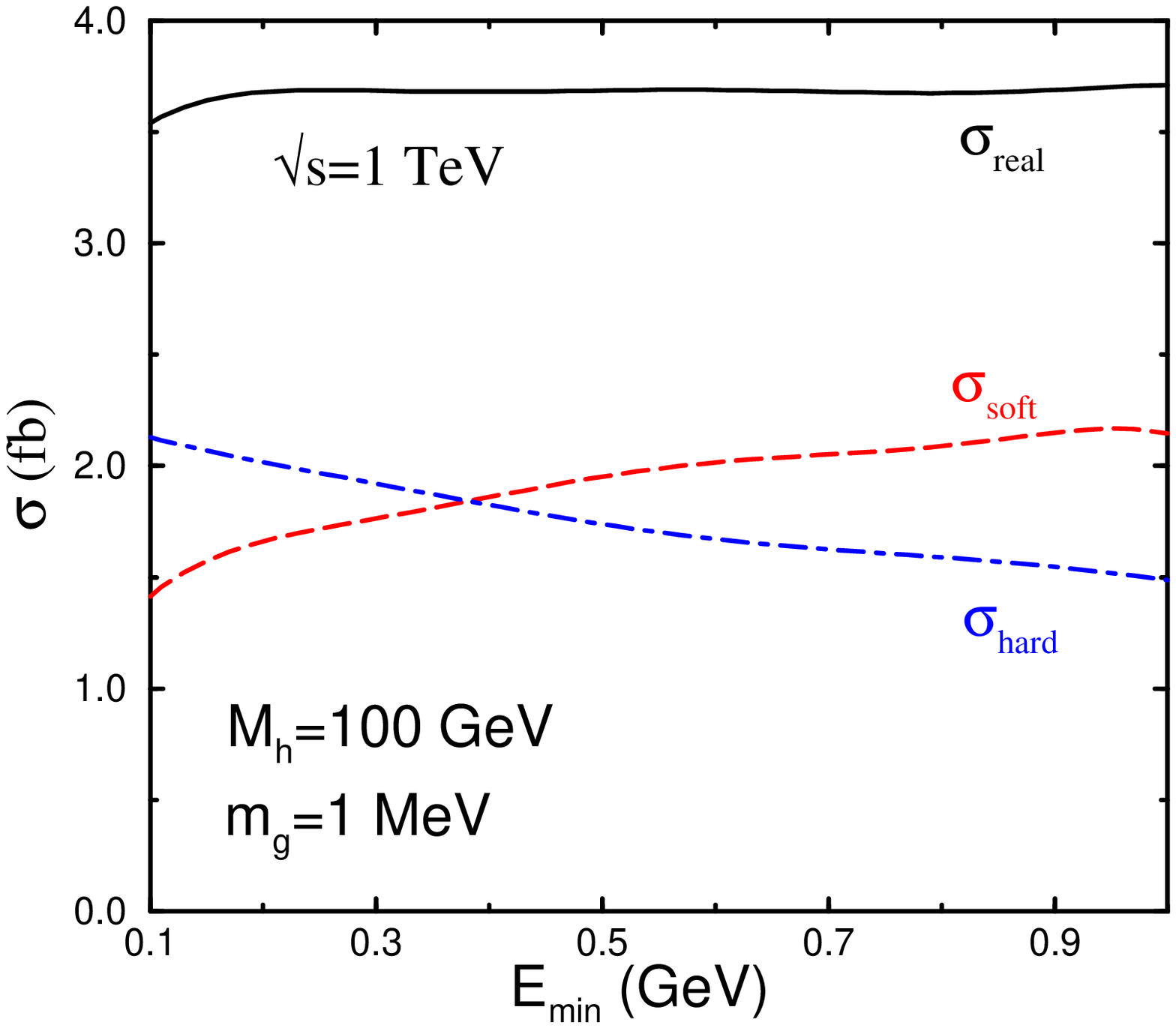}
\caption[]{Contributions from the process $e^+e^- \rightarrow
t {\overline t}h g$ as a function of the cut-off on the soft gluon
energy, $E_{min}$, for fixed gluon and Higgs boson masses at
$\sqrt{s}=1$~TeV.  We take $M_t=175$~GeV and $\alpha_s(M_t)=.11164$ }
\label{emin1000fig}
\end{figure} 

The integral over the soft gluon phase space has been performed in
Refs.~\cite{tv} and \cite{den} leading to the analytic result for the
soft contribution to $e^+e^-\rightarrow t {\overline t}h g$,
 
\bea
\biggl({d\sigma\over d x_h}\biggr)_{soft}&=&
\biggl({d\sigma\over d x_h}\biggr)_0 {\alpha_s C_F\over 2 
\pi}
\biggl\{
-2\log\biggl({4 E_{min}^2\over m_g^2}\biggr)
\biggl[1-{x p_t\cdot p_{t^\prime}\over
M_t^2(x^2-1)}\log(x^2
)\biggr]
\nonumber \\
&&
-{p_t^0\over \mid {\vec p_t}\mid }\log\biggl({p_t^0-\mid 
{\vec p_t}\mid
\over p_t^0+\mid {\vec p_t}\mid}\biggr)
-{p_{t^\prime}^0\over \mid {\vec p_{t^\prime}}
\mid }\log\biggl({p_{t^\prime}^0-\mid {\vec 
p_{t^\prime}}\mid
\over p_{t^\prime}^0+\mid {\vec p_{t^\prime}}\mid}\biggr)
\nonumber \\
&&
+{4x p_t\cdot p_{t^\prime}\over M_t^2(x^2-1)}
\biggl[
{1\over 4}\log^2\biggl({u^0-\mid{\vec u}\mid \over
u^0+\mid{\vec u}\mid}\biggr)
\nonumber \\
&& +Li_2\biggl(
1-{u^0+\mid {\vec u}\mid\over v}\biggr) 
+Li_2\biggl(
1-{u^0-\mid {\vec u}\mid\over v}\biggr)\biggr]^{u=x p_t}_
{u=p_{t^\prime}}
\biggr\}\,\,\,,
\label{softglu}
\eea 

\noindent where $x$ is the solution to

\be
M_t^2(x^2+1)-2x p_t\cdot p_{t^\prime} =0\,\,\,,
\ee 

\noindent subject to the constraint

\be
{x p_t^0- p_{t^\prime}^0\over p_{t^\prime}^0} > 0\,\,\,, 
\ee

\noindent and

\be
v={M_t^2(x^2-1)\over 2 (x p_t^0-p_{t^\prime}^0)}\,\,\,.
\ee 

\subsection{Hard gluon radiation}
\label{sec:realhard}

The hard gluon contribution is calculated from the diagrams of
Fig. \ref{glufig} for gluon momenta $E_g>E_{min}$. The integration
over the final state phase space is done using a numerical Monte
Carlo.  The hard gluon cross section is insensitive to the small value
chosen for the gluon mass.  In Figs. \ref{eminfig} and
\ref{emin1000fig} we show the contributions of the soft and hard 
gluon radiation as a function of $E_{min}$.  The sum of the soft and
hard terms is clearly independent of the separation, $E_{min}$ for
$E_{min}>>m_g$.  The result retains a dependence on the gluon mass,
$m_g$, however, which is cancelled when the infrared divergent pieces
of the virtual contributions are included (see
Section~\ref{sec:virtual}).

We choose $v=(\sqrt{2}G_F)^{-1/2}\simeq 246$~GeV, $M_t=175$~GeV, and
use the 1-loop value for $\alpha_s(M_t)=.11164$ (which corresponds to
$\Lambda_{QCD}=200$~MeV) in all our numerical calculations.

\section{Virtual Corrections}
\label{sec:virtual}

The amplitude for $e^+e^-\rightarrow t\bar t h$ including virtual
corrections to $O(\alpha_s)$ can be written as

\be
{\cal A} = {\cal A}_0 + \frac{\alpha_s}{4\pi}C_F {\cal 
A}_1^{\rm virt}
\,\,\,,
\ee

\noindent where $C_F=4/3$. We include in ${\cal A}_1^{\rm 
virt}$ also the wave function, Yukawa coupling, and mass
counterterms. The corresponding contribution to the cross section at
$O(\alpha_s)$ is

\be 
\sigma^{\rm virt} = \frac{\alpha_s}{4\pi}C_F\,
2\real({\cal A}_1^{\rm virt}{\cal A}_0^*) = 
\frac{\alpha_s}{2\pi}C_F\real({\cal A}_1^{\rm virt}{\cal
  A}_0^*)\,\,\,.  
\ee

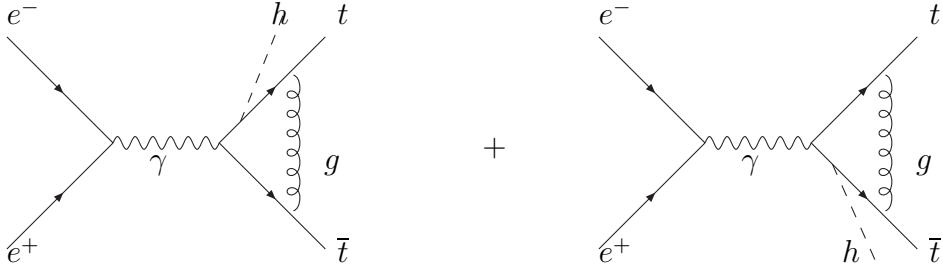
\begin{figure}[btp]
\begin{picture}(100,100)(-20,-5)
\SetScale{0.8}
\ArrowLine(0,100)(50,50)
\ArrowLine(0,0)(50,50)
\Photon(50,50)(100,50){3}{6}
\ArrowLine(100,50)(150,100) 
\ArrowLine(100,50)(150,0)
\DashLine(110,60)(130,110){5}
\Gluon(135,82)(135,18){3}{6} 
\put(180,35){$+$} 
\put(54,30){$\gamma$} 
\put(0,85){$e^-$}
\put(0,-5){$e^+$}
\put(100,85){$h$}
\put(120,30){$g$} 
\put(125,85){$t$}
\put(125,-5){${\overline t}$}
\SetScale{1}
\end{picture}
\begin{picture}(100,100)(-140,-5)
\SetScale{0.8}
\ArrowLine(0,100)(50,50)
\ArrowLine(0,0)(50,50)
\Photon(50,50)(100,50){3}{6}
\ArrowLine(100,50)(150,100) 
\ArrowLine(100,50)(150,0)
\Gluon(135,82)(135,18){3}{6}
\DashLine(110,40)(130,-5){5}
\put(54,30){$\gamma$} 
\put(0,85){$e^-$}
\put(0,-5){$e^+$}
\put(92,-5){$h$}
\put(120,30){$g$} 
\put(125,85){$t$}
\put(125,-5){${\overline t}$}
\SetScale{1}
\end{picture}  
\caption[]{Box diagrams contributing to the virtual 
corrections to $e^+e^-\rightarrow t {\overline t} h$.}
\label{boxfeyndiag} 
\end{figure}

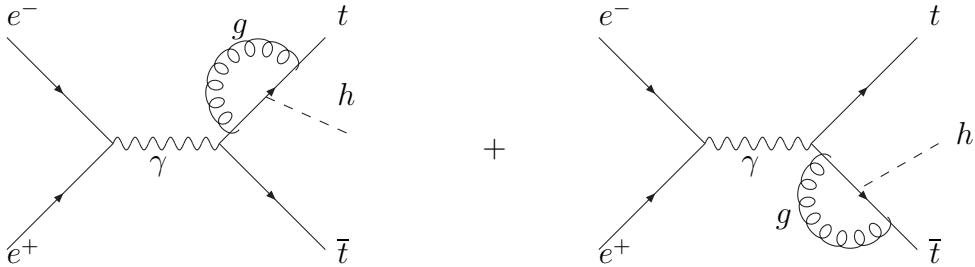
\begin{figure}[hbtp]
\begin{picture}(100,100)(-20,-5)
\SetScale{0.8}
\ArrowLine(0,100)(50,50)
\ArrowLine(0,0)(50,50)
\Photon(50,50)(100,50){3}{6}
\ArrowLine(100,50)(150,100) 
\ArrowLine(100,50)(150,0)
\DashLine(122,72)(160,55){5}
\GlueArc(118,75)(20.355,28,245){4}{8} 
\put(180,35){$+$} 
\put(54,30){$\gamma$} 
\put(0,85){$e^-$}
\put(0,-5){$e^+$}
\put(125,55){$h$}
\put(85,82){$g$} 
\put(125,85){$t$}
\put(125,-5){${\overline t}$}
\SetScale{1}
\end{picture}
\begin{picture}(100,100)(-140,-5)
\SetScale{0.8}
\ArrowLine(0,100)(50,50)
\ArrowLine(0,0)(50,50)
\Photon(50,50)(100,50){3}{6}
\ArrowLine(100,50)(150,100) 
\ArrowLine(100,50)(150,0)
\GlueArc(118,25)(20.355,-245,-28){4}{8}
\DashLine(125,30)(160,50){5}
\put(54,30){$\gamma$} 
\put(0,85){$e^-$}
\put(0,-5){$e^+$}
\put(135,40){$h$}
\put(67,10){$g$} 
\put(125,85){$t$}
\put(125,-5){${\overline t}$}
\SetScale{1}
\end{picture}
\caption[]{Vertex diagrams of \emph{type 1} contributing 
to the virtual corrections to $e^+e^-\rightarrow t {\overline t} h$.}
\label{virtfig1} 
\end{figure}

\begin{figure}[btp]
\begin{picture}(100,100)(-20,-5)
\SetScale{0.8}
\ArrowLine(0,100)(50,50)
\ArrowLine(0,0)(50,50)
\Photon(50,50)(100,50){3}{6}
\ArrowLine(100,50)(150,100) 
\ArrowLine(100,50)(150,0)
\DashLine(140,90)(160,55){5}
\Gluon(135,82)(135,18){3}{6} 
\put(180,35){$+$} 
\put(54,30){$\gamma$} 
\put(0,85){$e^-$}
\put(0,-5){$e^+$}
\put(125,55){$h$}
\put(97,40){$g$} 
\put(125,85){$t$}
\put(125,-5){${\overline t}$}
\SetScale{1}
\end{picture}
\begin{picture}(100,100)(-140,-5)
\SetScale{0.8}
\ArrowLine(0,100)(50,50)
\ArrowLine(0,0)(50,50)
\Photon(50,50)(100,50){3}{6}
\ArrowLine(100,50)(150,100) 
\ArrowLine(100,50)(150,0)
\Gluon(135,82)(135,18){3}{6}
\DashLine(140,10)(160,50){5}
\put(54,30){$\gamma$} 
\put(0,85){$e^-$}
\put(0,-5){$e^+$}
\put(135,40){$h$}
\put(97,40){$g$} 
\put(125,85){$t$}
\put(125,-5){${\overline t}$}
\SetScale{1}
\end{picture}
\caption[]{Vertex diagrams of \emph{type 2} contributing 
to the virtual corrections to $e^+e^-\rightarrow t {\overline t} h$.}
\label{virtfeyn}
\end{figure}
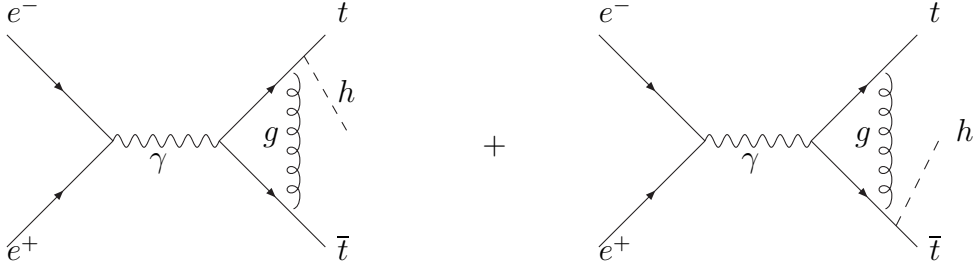

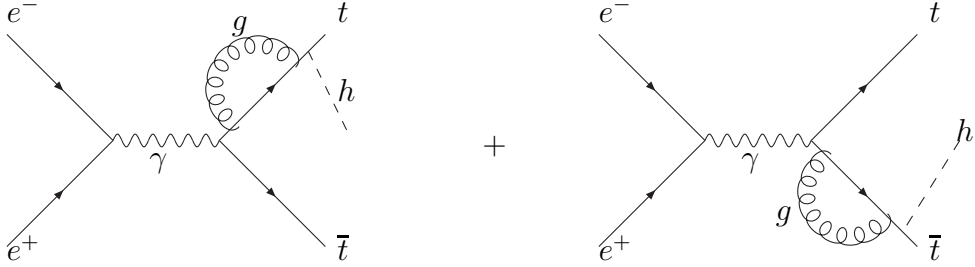
\begin{figure}[hbtp]
\begin{picture}(100,100)(-20,-5)
\SetScale{0.8}
\ArrowLine(0,100)(50,50)
\ArrowLine(0,0)(50,50)
\Photon(50,50)(100,50){3}{6}
\ArrowLine(100,50)(150,100) 
\ArrowLine(100,50)(150,0)
\DashLine(142,92)(160,55){5}
\GlueArc(118,75)(20.355,28,245){4}{8} 
\put(180,35){$+$} 
\put(54,30){$\gamma$} 
\put(0,85){$e^-$}
\put(0,-5){$e^+$}
\put(125,55){$h$}
\put(85,82){$g$} 
\put(125,85){$t$}
\put(125,-5){${\overline t}$}
\SetScale{1}
\end{picture}
\begin{picture}(100,100)(-140,-5)
\SetScale{0.8}
\ArrowLine(0,100)(50,50)
\ArrowLine(0,0)(50,50)
\Photon(50,50)(100,50){3}{6}
\ArrowLine(100,50)(150,100) 
\ArrowLine(100,50)(150,0)
\GlueArc(118,25)(20.355,-245,-28){4}{8}
\DashLine(145,10)(170,50){5}
\put(54,30){$\gamma$} 
\put(0,85){$e^-$}
\put(0,-5){$e^+$}
\put(135,40){$h$}
\put(67,10){$g$} 
\put(125,85){$t$}
\put(125,-5){${\overline t}$}
\SetScale{1}
\end{picture}
\caption[]{Feynman diagrams contributing to the self 
energy corrections of the internal heavy quark propagators for
$e^+e^-\rightarrow t {\overline t} h$.}
\label{self} 
\end{figure} 

\begin{figure}[tb]
\centering
\epsfxsize=4.5in
\leavevmode\epsffile{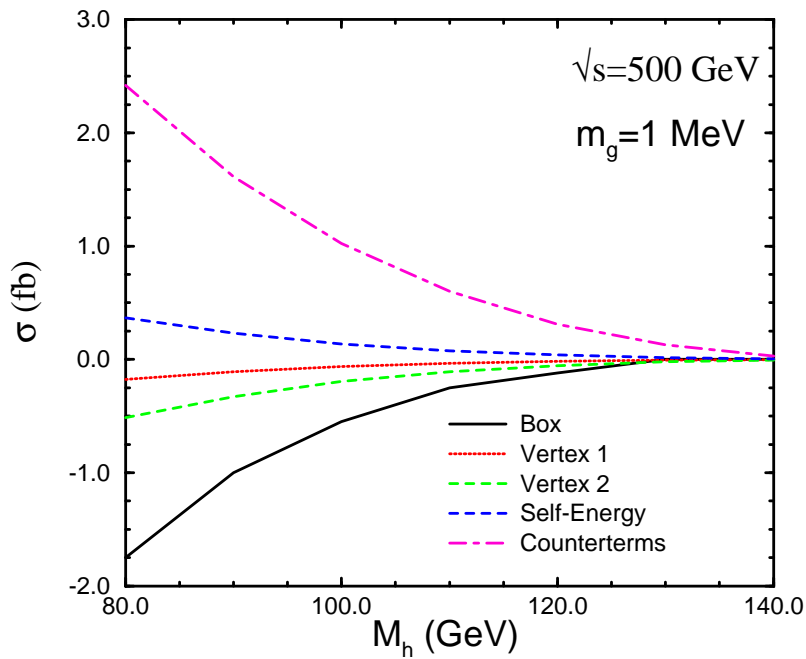}
\caption[]{Contributions to $e^+e^-\rightarrow t 
{\overline t} h$ from virtual diagrams at $\sqrt{s}=500$~GeV with a
finite gluon mass.  We take $M_t=175$~GeV and $\alpha_s(M_t)=.11164$}
\label{virtcons}
\end{figure}

\noindent
The virtual corrections we consider are:
\begin{itemize}
\item box diagrams (2 diagrams, see 
Fig.~\ref{boxfeyndiag});
\item vertex corrections of \emph{type 1} (2 diagrams, 
see Fig.~\ref{virtfig1} );
\item vertex corrections of \emph{type 2} (2 diagrams, 
see Fig.~\ref{virtfeyn});
\item self-energy corrections to the internal leg 
propagators (2
diagrams, see Fig.~\ref{self});
\item wave function and Yukawa coupling renormalization;
\item mass renormalization (2 counterterm insertions, one 
for each
internal propagator).
\end{itemize}

\noindent In general, virtual diagrams can have both 
ultraviolet (UV) and infrared (IR) singularities. In order to
explicitly check the cancellation of both UV and IR divergences among
different loop diagrams and counterterms, we have regularized them
using dimensional regularization.  We have then verified the
cancellation of the $1/\epsilon_{UV}$ and $1/\epsilon_{IR}$ poles both
analytically and numerically. We will discuss IR divergences in
greater detail in Section~\ref{sec:irdiv}.
 
\subsection{Wave Function and Mass Renormalization 
Counterterms}
\label{sec:virtualcount}

The wave function renormalization contribution to the cross section
can be expressed directly in terms of $\sigma_0$, the cross section at
the tree level for $e^+e^-\rightarrow t {\overline t} h$,

\be
\sigma_{Z_2}=\sigma_0^\epsilon \biggl( 2 \delta 
Z_2\biggr)\,\,\,.
\ee

\noindent The tree level cross section must be computed 
to ${\cal O}(\epsilon)$, hence we denote it by $\sigma_0^\epsilon$.
We separate the infrared and ultraviolet divergences and write the
wave function renormalization countertern as\footnote{We thank the
authors of Ref.~[4] for pointing out an inconsistency in our original
treatment of the wavefunction renormalization.},

\bea 
Z_2 &=& 1+\delta Z_2 \\
&=&1 -\frac{\alpha_s}{4\pi}C_F\biggl({4\pi\mu^2
\over M_t^2}\biggr)^\epsilon\Gamma(1+\epsilon)
\left(
\frac{1}{\epsilon_{UV}}+\frac{2}{\epsilon_{IR}}
+4\right)\,\,\,.\nonumber
\label{z2def} 
\eea

\noindent The renormalization of the Yukawa coupling 
constant contributes to the ${\cal O}(\alpha_s)$ cross section,

\be
\sigma_{Y}=\sigma_0^\epsilon
\biggl({\delta M_t\over M_t}\biggr)\,\,\,,
\ee

\noindent where we use the pole definition of the top 
quark mass,

\be
{\delta M_t\over M_t}=-{\alpha_s\over 4\pi} C_F
\biggl({4\pi\mu^2\over 
M_t^2}\biggr)^\epsilon\Gamma(1+\epsilon)
 \biggl(
{3\over \epsilon_{UV}}+4\biggr)\,\,\,. 
\label{masren} 
\ee

Finally, there is the mass renormalization of the internal heavy quark
propagators which is calculated using Eq.~(\ref{masren}).

\subsection{Calculation of Loop Diagrams}

The analytical reduction of the one loop diagrams to a linear
combination of one loop tensor and scalar integrals is performed using
FORM, MAPLE and MAXIMA. The one loop integrals are then evaluated with
the help of the numerical program FF\cite{ff}.  For a fixed set of
momenta, this program evaluates the finite contribution to the loop
integrals numerically. The integrals over the phase space are
performed with a numerical Monte Carlo.  In Fig.
\ref{virtcons}, we show the finite contributions from the virtual
diagrams.  The virtual contribution can be written as

\be
\sigma_{virt}=\sigma_{box}+\sigma_{V1}+\sigma_{V2}+\sigma
_{self}+
\sigma_{count}\quad ,
\ee

\noindent where $\sigma_{box}, \sigma_{V1},\sigma_{V2}$ and 
$\sigma_{self}$ are shown in Figs. \ref{boxfeyndiag}~-~\ref{self}.
The counterterms are $\sigma_{count}=
\sigma_{Z_2}+\sigma_Y+\sigma_m$, where $\sigma_m$ is the 
contribution of the internal propagator mass renormalization.
Individually, the diagrams generate $\log(\mu^2/M_t^2)$ contributions
which are associated with $1/\epsilon_{UV}$ singularities and we take
the renormalization scale $\mu=M_t$. Some virtual corrections also
generate IR divergences and we discuss these in
Section~\ref{sec:irdiv}. The large cancellations between the various
contributions is clear from Fig.  ~\ref{virtcons}.

\subsection{Cancellation of Infrared Divergences}
\label{sec:irdiv}

\begin{figure}[htb]
\centering
\epsfxsize=4.5in
\leavevmode\epsffile{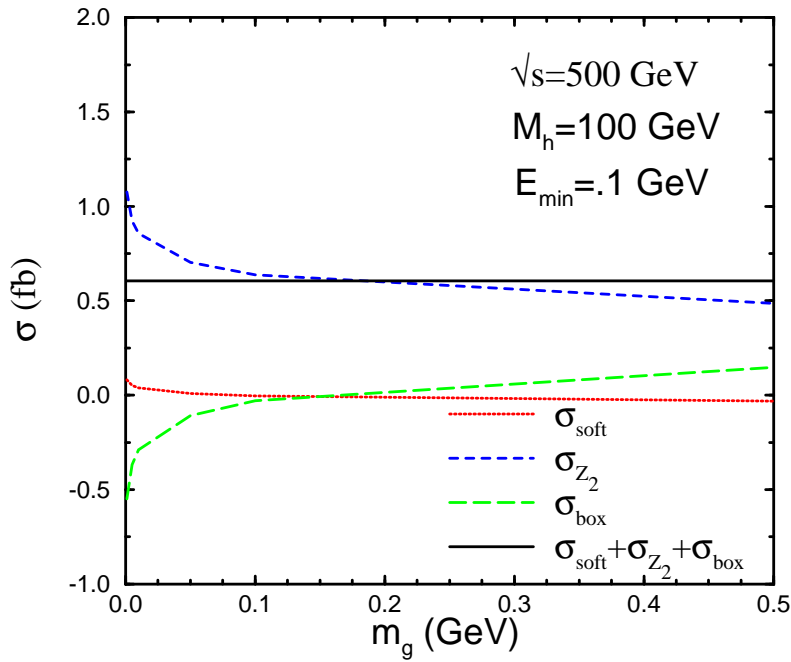}
\caption[]{Contributions to $e^+e^-\rightarrow t 
{\overline t} h$ which have infrared divergences.  We take
$M_t=175$~GeV and $\alpha_s(M_t)=.11164$}
\label{mglu500fig}
\end{figure}

\begin{figure}[htb]
\centering
\epsfxsize=4.5in
\leavevmode\epsffile{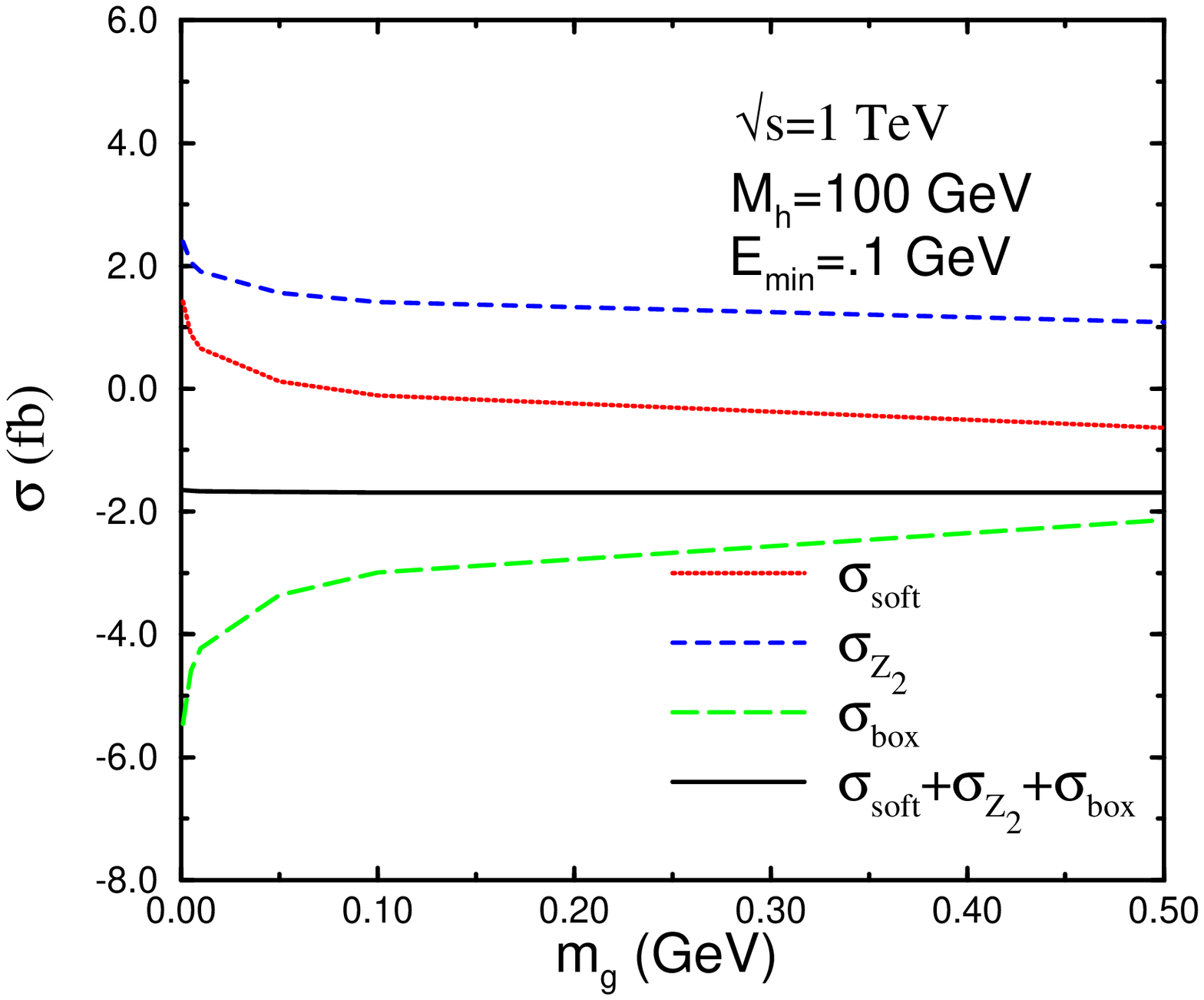}
\caption[]{Contributions to $e^+e^-\rightarrow t 
{\overline t} h$ which have infrared divergences.  We take
$M_t=175$~GeV and $\alpha_s(M_t)=.11164$}
\label{mglu1000fig}
\end{figure}

As we discussed before, we have verified the cancellation of the
$1/\epsilon_{UV}$ ultraviolet singularities both analytically and
numerically. The cancellation of the $1/\epsilon_{IR}$ singularities
has also been verified analytically. Equivalently, the infrared
divergences can be regulated by introducing a finite gluon mass, as
was done for the real gluon diagrams of Section~\ref{sec:real}. This
amounts to making the substitution

\be
\Gamma(\epsilon)(4\pi\mu^2)^\epsilon\rightarrow 
\log(m_g^2)\,\,\,
\ee 

\noindent which is what we have used in our  numerical 
calculations.

Infrared divergences arise from the box diagram, Fig.
\ref{boxfeyndiag}, the $Z_2$ wavefunction renormalization, 
Eq.~(\ref{z2def}), and the soft gluon emission, Eq.~(\ref{softglu}).
In Figs.~\ref{mglu500fig} and \ref{mglu1000fig} we show the sum of
these contributions as a function of the gluon mass and see that the
sum is independent of the gluon mass, confirming the cancellation of
the infrared divergences.

\section{Results}
\label{sec:results}

\begin{figure}[htb]
\centering
\epsfxsize=4.5in
\leavevmode\epsffile{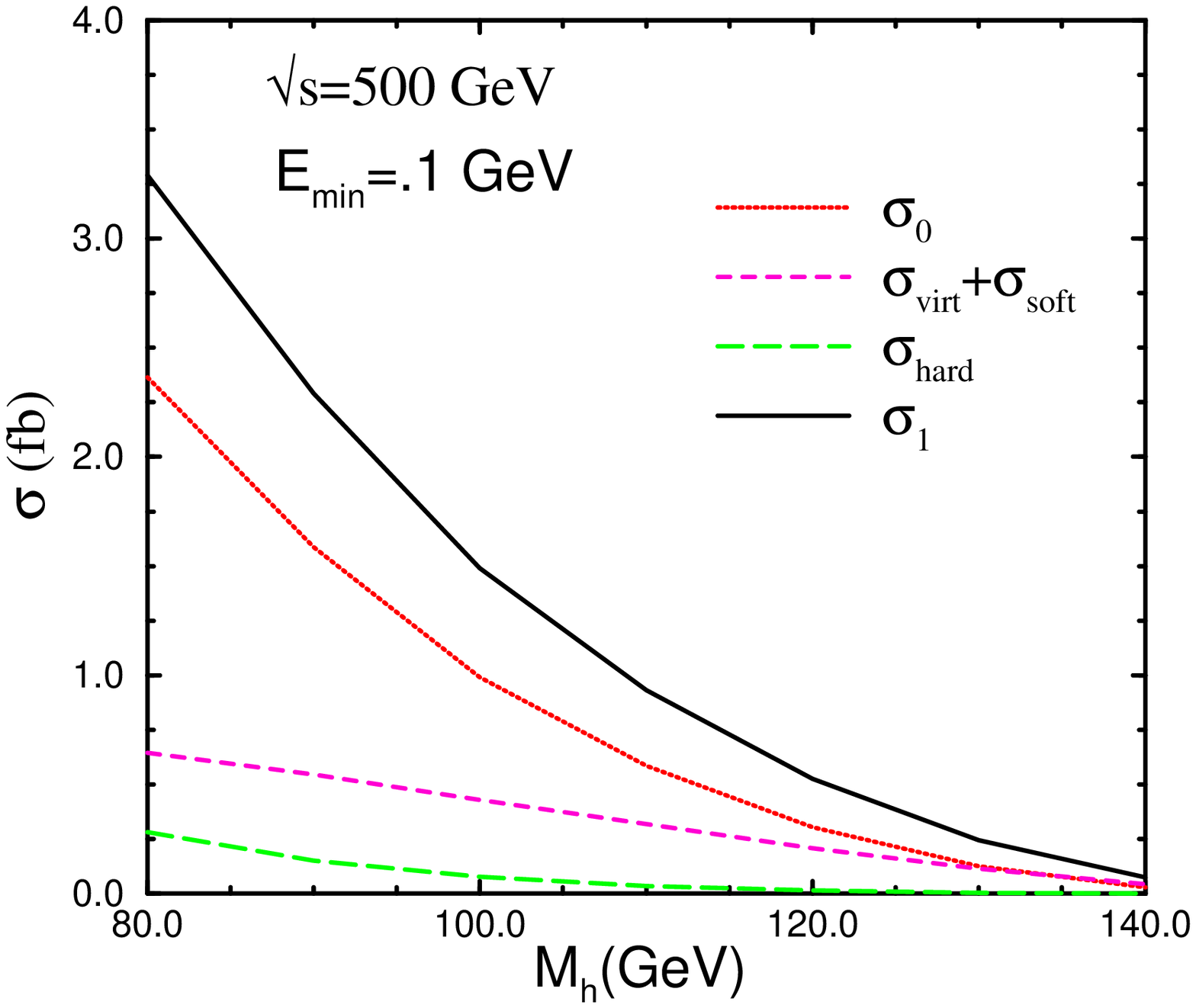}
\caption[]{QCD corrections to $e^+e^-\rightarrow 
t {\overline t}h$ at $\sqrt{s}=500$~GeV.  $\sigma_0$ is the lowest
order cross section and $\sigma_1$ is the complete ${\cal
O}(\alpha_s)$ corrected rate.  We take $M_t=175$~GeV and
$\alpha_s(M_t)=.11164$}
\label{sig500fig}
\end{figure}

\begin{figure}[htb]
\centering
\epsfxsize=4.5in
\leavevmode\epsffile{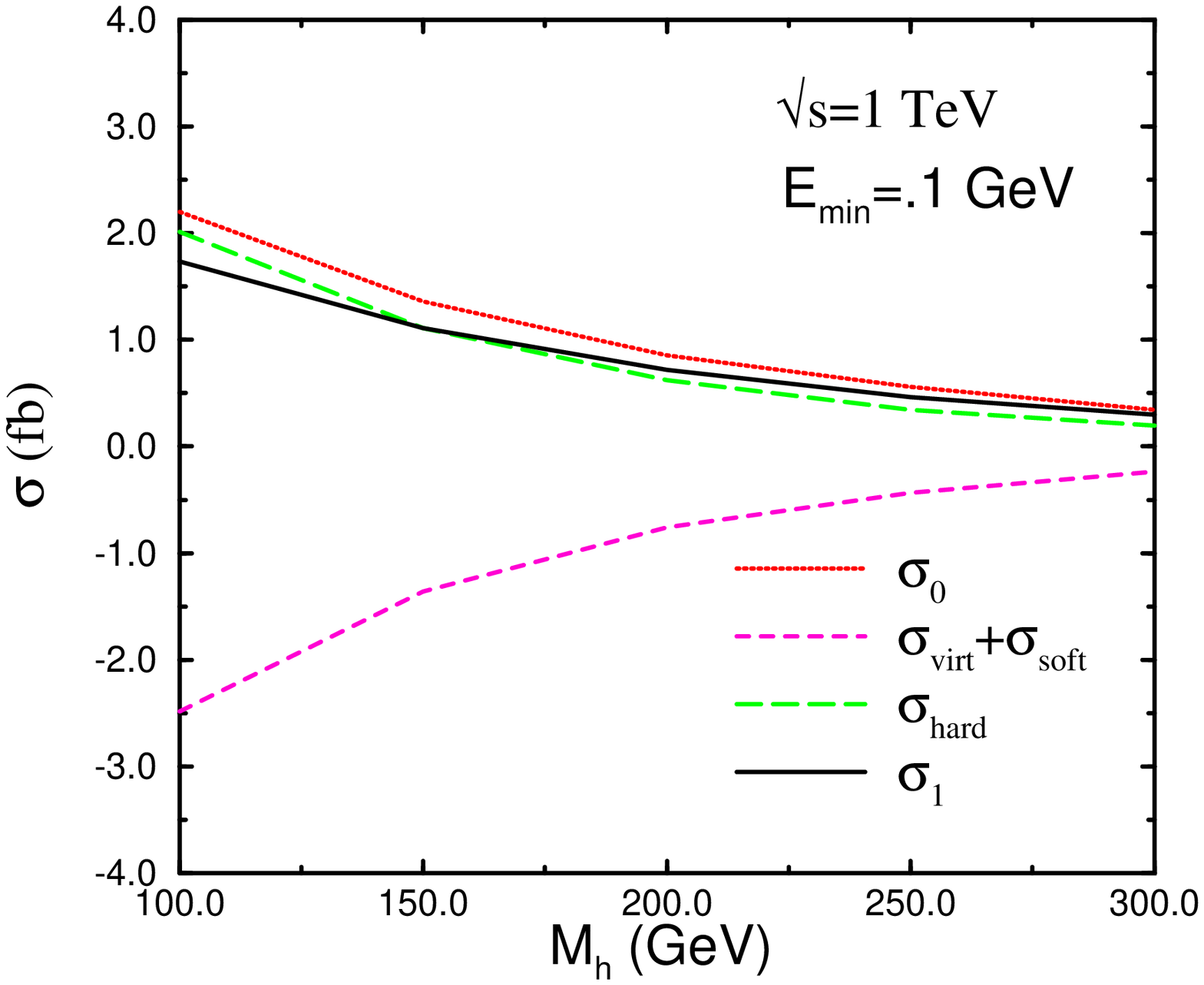}
\caption[]{QCD corrections to $e^+e^-\rightarrow 
t {\overline t}h$ at $\sqrt{s}=1$~TeV.  $\sigma_0$ is the lowest order
cross section and $\sigma_1$ is the complete ${\cal O}(\alpha_s)$
corrected rate.  We take $M_t=175$~GeV and $\alpha_s(M_t)=.11164$}
\label{sig1000fig}
\end{figure}

\begin{figure}[htb]
\centering
\epsfxsize=4.5in
\leavevmode\epsffile{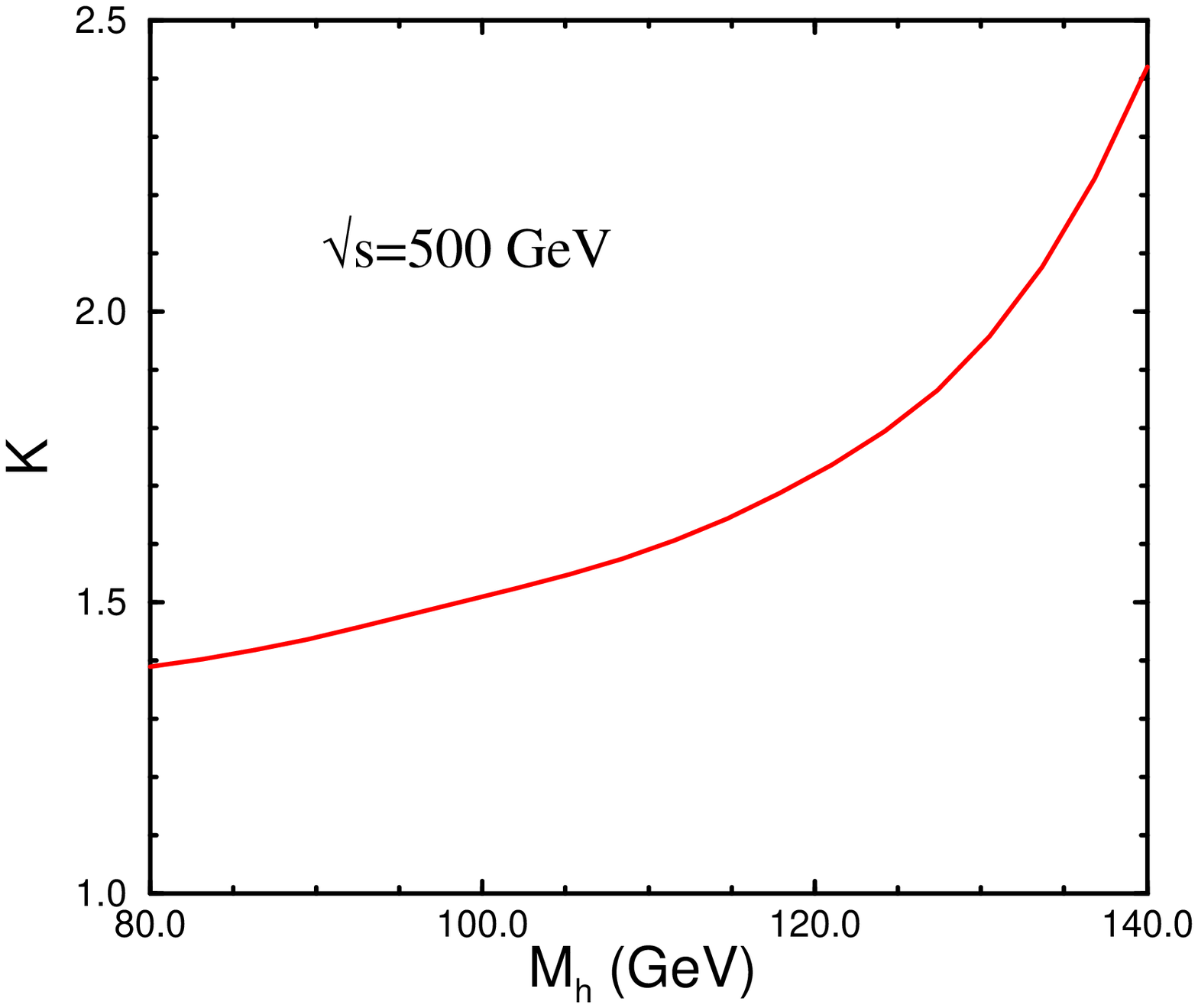}
\caption[]{Ratio of the ${\cal O}(\alpha_s)$ corrected 
rate to the lowest order cross section for $e^+e^-\rightarrow t
{\overline t}h$ at $\sqrt{s}=500$~GeV.  We take $M_t=175$~GeV and
$\alpha_s(M_t)=.11164$}
\label{kfac500fig}
\end{figure}

\begin{figure}[htb]
\centering
\epsfxsize=4.5in
\leavevmode\epsffile{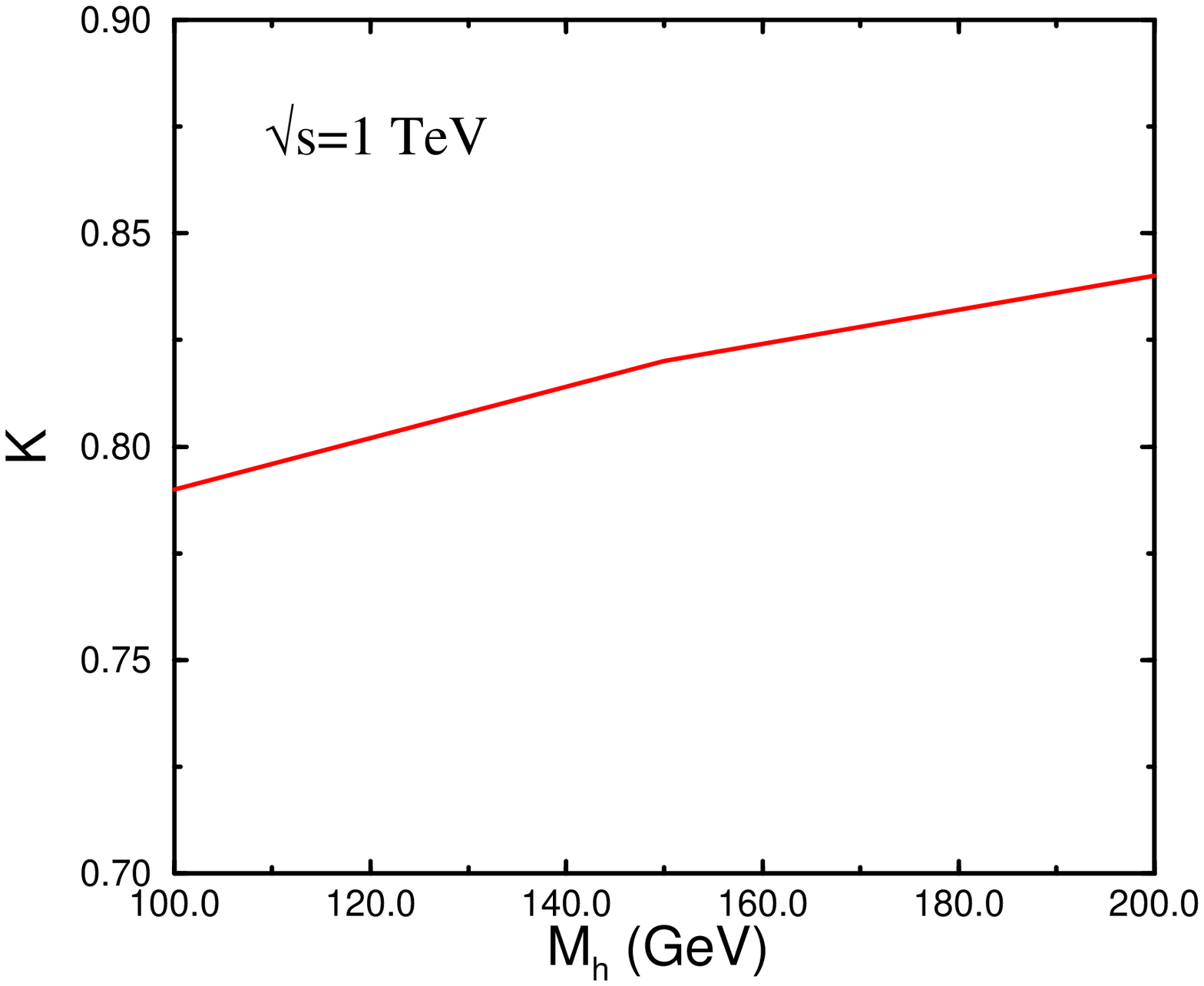}
\caption[]{Ratio of the ${\cal O}(\alpha_s)$ corrected 
rate to the lowest order cross section for $e^+e^-\rightarrow t
{\overline t}h$ at $\sqrt{s}=1$~TeV.  We take $M_t=175$~GeV and
$\alpha_s(M_t)=.11164$}
\label{kfac1000fig}
\end{figure} 

\subsection{Numerical Results}
\label{sec:numerics}

When the virtual and real contributions are combined, the final result
is finite and independent of both $E_{min}$ and $m_g$.  In
Figs.~\ref{sig500fig} and \ref{sig1000fig}, we show the various
contributions to the total cross section.  $\sigma_1$ is the complete
${\cal O}(\alpha_s)$ corrected rate,

\be
\sigma_1=\sigma_0+\sigma_{virt}+\sigma_{hard}+\sigma_{sof
t}\,\,\,.
\ee

\noindent The combination $\sigma_{virt}+\sigma_{soft}$ 
is independent of the gluon mass, but retains a dependence on
$E_{min}$ which is cancelled by $\sigma_{hard}$.  At
$\sqrt{s}=500$~GeV, the corrections are large and positive,
significantly increasing the rate.  The corrections are smaller at
$\sqrt{s}=1$~TeV, with large cancellations between the hard and the
virtual plus soft contributions.

The size of the QCD corrections can be described by a $K$ factor,

\be
K(\mu)\equiv{\sigma_1\over \sigma_0}\,\,\,,
\label{kdef}
\ee

\noindent 
which is shown in Figs.~ \ref{kfac500fig} and ~\ref{kfac1000fig}.
Note that after the cancellation of the $1/\epsilon_{UV}$ divergences,
the only $\mu$ dependence is in $\alpha_s(\mu)$.  If $\mu=\sqrt{s}$,
then $K(M_h=100\,\mbox{GeV})$ is reduced to $1.4$ from the value
$K=1.5$ obtained with $\mu=M_t$ for $\sqrt{s}=500$~GeV.  The authors
of Ref.~\cite{dittqcd} choose $\mu=\sqrt{s}$.  Our results for the $K$
factor are in agreement with theirs.

\begin{figure}[tb]
\centering
\epsfxsize=4.5in
\leavevmode\epsffile{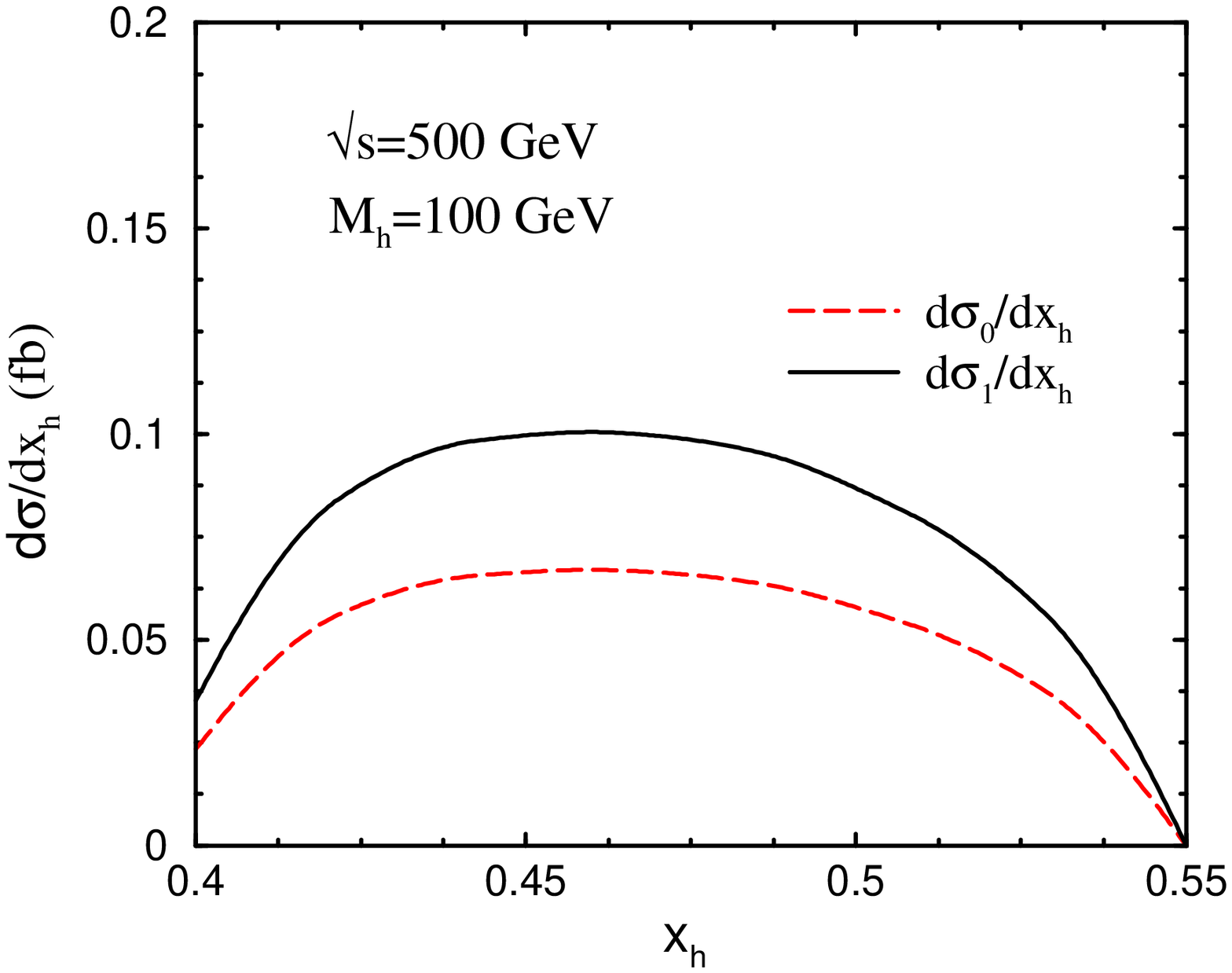} 
\caption[]{Differential cross section, $d\sigma/dx_h$,
for $e^+e^-\rightarrow t {\overline t}h$ for $M_h=100$~GeV and
$\sqrt{s}=500$~GeV.  $\sigma_0$ is the lowest order cross section and
$\sigma_1$ is the complete ${\cal O}(\alpha_s)$ corrected rate.  We
take $M_t=175$~GeV and $\alpha_s(M_t)=.11164$.}
\label{dsigfig}
\end{figure}

In Fig. \ref{dsigfig}, we show the shape of the differential cross
section $d\sigma\over d x_h$ at $\sqrt{s}=500$~GeV and for
$M_h=100$~GeV.  The cross section is peaked around $x_h\sim .45$.
Including the ${\cal O}(\alpha_s)$ corrections has little effect on
the shape of the distribution.

\subsection{Comparison with Approximate Result}
\label{sec:eha}

In Ref.~\cite{dr}, the ${\cal O}(\alpha_s)$ corrections to the process
$e^+e^-\rightarrow t {\overline t}h$ are computed in a framework where
the Higgs boson is treated as a parton which is radiated from the
heavy top quark.  The distribution of Higgs bosons in a top quark,
$f_{t\rightarrow h}(x_h)$ is computed to ${\cal O}(\alpha_s)$ and
convoluted with the cross section for $e^+ e^-\rightarrow t {\overline
t}$, also computed to ${\cal O}(\alpha_s)$,

\be 
\sigma(e^+e^-\rightarrow t {\overline t}h)_{EHA} =
2\int^{x_h^{max}}_{x_h^{min}} dx_h f_{t\rightarrow 
h}(x_h)
\sigma(e^+e^-\rightarrow t {\overline t})\,\,\,.  
\label{ehadef}
\ee 

\noindent This approximation, which we call the Effective Higgs
Approximation (EHA), is expected to be valid for $M_h << M_t
<<\sqrt{s}$.

The impact of QCD corrections on the prediction of the cross section
for $e^+e^-\rightarrow t\bar t h$ is described by the K-factor of
Eq.~(\ref{kdef}), where the cross sections are evaluated using
Eq.~(\ref{ehadef}) to the appropriate order in $\alpha_s$.  In
Ref.~\cite{dr}, we found $K(\mu=\sqrt{s})\!\simeq .94$ at
$\sqrt{s}=1$~TeV for $M_h\sim 100$~GeV.  Comparing with
Fig.~\ref{kfac1000fig}, we see good agreement between the EHA at
$\sqrt{s}=1$~TeV and the exact calculation presented here.  However,
the hard gluon bremsstrahlung terms of Section~\ref{sec:real} cannot be
adequately included within the context of the EHA. From
Fig.~\ref{emin1000fig}, it is apparent that the hard gluon terms are
not small. Hence the agreement between the EHA and the present
calculation seems to derive from some more complicated cancellation
between the hard gluon terms and those $O(M_t^2/s)$, $O(M_h^2/s)$ and
$O(M_h^2/M_t^2)$ contributions that are also sistematically neglected
in the EHA.

\section{Conclusion} 
\label{sec:conclusions}

We have computed the ${\cal O}(\alpha_s)$ corrected rate for
$e^+e^-\rightarrow t {\overline t}h$.  At $\sqrt{s}=500$~GeV, the
corrections are large and positive, while at $\sqrt{s}=1$~TeV, the QCD
corrections are small.  Studies of the experimental viability of this
process as a means of measuring the top quark -Higgs boson Yukawa
coupling are needed in order to assess the usefulness of the $t
{\overline t}h$ process.

\section*{Acknowledgments}
We are grateful for the generous hospitality of the I.C.T.P. and the
S.I.S.S.A. institutes while this work was being completed.  We thank
S.~Dittmaier, M.~Kramer, Y. Li, M.~Spira, and P.~Zerwas for discussion
of their results prior to publication.  We also thank A.~Czarnecki for
helpful discussions and G.~J.~van~Oldenborgh for suggestions about his
code FF.  The work of S.~D.  is supported by the U.S. Department of
Energy under contract DE-AC02-76CH00016. The work of L.~R. is
supported by the U.S.  Department of Energy under contract
DE-FG02-95ER40896.

\end{document}